\documentstyle[epsfig]{aipproc}

\def\la{\ifmmode\stackrel{<}{_{\sim}}\else$\stackrel{<}{_{\sim}}$\fi} 
\def\ga{\ifmmode\stackrel{>}{_{\sim}}\else$\stackrel{>}{_{\sim}}$\fi}

\begin{document}
\title{Plerions and Pulsar-Powered Nebulae}

\author{Bryan M. Gaensler\thanks{Hubble Fellow}}
\address{Center for Space Research, Massachusetts Institute of
Technology, Cambridge MA 02139, USA}

%\lefthead{LEFT head}
%\righthead{RIGHT head}
\maketitle

\begin{abstract}

In this brief review, I discuss recent developments in
the study of pulsar-powered nebulae (``plerions''). The large
volume of data which has been acquired in recent years reveals
a diverse range of observational properties, demonstrating
how differing environmental and pulsar properties
manifest themselves in the resulting nebulae.

\end{abstract}

\section*{Introduction}

All isolated pulsars have steadily increasing periods. However, only
a very small fraction of the
corresponding spin-down luminosity, 
$\dot{E} \equiv 4\pi^2 I\dot{P}/P^3$ (where $P$ is
the pulsar spin period and $I$ is the pulsar's moment of inertia),
can usually be
accounted for by the pulsations themselves. It is rather
assumed that most of this rotational energy is converted
into a relativistic wind. 

At some distance from the pulsar,
the wind pressure and confining pressure
balance, producing a shock at which relativistic particles
in the wind are accelerated and have their pitch angles
randomized. These particles consequently radiate synchrotron
emission, generating an observable pulsar-powered nebula, or
``plerion''.\footnote{For simplicity, I will use the term ``plerion''
to refer generically to all forms of pulsar nebulae,
and will make specific distinctions when I refer to particular
sub-categories of object.} 
Plerions can be powerful
probes of a pulsar's interaction with its surroundings ---
their properties can be used to infer the geometry,
energetics and composition of the pulsar wind, the space velocity of
the pulsar itself, and the properties of the ambient medium.
Furthermore, the mere existence of a plerion indicates the presence
of a pulsar within, even when the latter has not yet
been directly detected.

In the following discussion, I briefly review our historical and theoretical
understanding of plerions, and describe some of the highlights
from recent work. There have been
many recent developments in this field, and space
precludes me from providing a more comprehensive review.
I refer the reader to preceding reviews on plerions and
their pulsars by Slane et al \cite{sbt98}, Chevalier
\cite{che98}, Frail \cite{fra98b,fra98}
and Gaensler \cite{gae01}.

\section*{Historical Overview}

Any discussion of plerions and pulsar wind nebulae must of course
begin with the Crab Nebula. While it has long been
realized that the Crab Nebula is the product
of a supernova explosion \cite{lun21}, its filled-center
morphology, flat radio spectrum and high fraction of linear
polarization all indicate that it is a very different 
source from ``shell-type'' supernova
remnants (SNRs) like Cassiopeia~A and the Cygnus Loop.
A pulsar was discovered embedded in the Crab Nebula in 1968 \cite{sr68}, and
it is the continual injection of energy into the nebula by this source
which is believed to produce the Crab's unusual properties.

In the early 1970s,
it was suggested that the filled-center SNR 3C~58 
was similarly powered by
an (as yet unseen) pulsar \cite{ws71}. Before long, several other such sources
were identified, and a whole class of ``Crab-like'' SNRs
began to emerge \cite{lgcm77,wp78,ws78}. 
Weiler \& Panagia \cite{wp78} proposed the name ``plerion'' 
for such sources, from the Greek ``$\pi\lambda\eta\rho\eta\varsigma$''
meaning ``full''.\footnote{Shakeshaft \cite{sha79}
has pointed out that ``$\pi\lambda\eta\rho\iota o \nu$''
is not a Greek word, and that ``plethoric supernova remnant'' would be
a more linguistically-correct term to describe these objects.
However, this advice does not seem to have been heeded by the
community!}

Milne et al \cite{mgh+79}
introduced an additional complication, when
they pointed out that the SNR~MSH~15--5{\em 6}, while having
a limb-brightened radio shell like most SNRs, also contains
a central core which otherwise resembles a plerion (see Figure~\ref{fig1}).
They termed this source a ``composite'' SNR, and proposed that
it combined the properties of shell-type SNRs and of plerions.

In the last 20 years, this simple classification of SNRs into shells,
plerions and composites has largely remained unchanged. In Green's most
recent Galactic SNR Catalog \cite{gre00}, 225 SNRs are listed, of which 23 are
composites and nine are plerions.  Of these 32 remnants which are presumed
to be powered by a pulsar, the central pulsars have now been
detected in ten, while a further five contain central X-ray point sources
believed to be associated neutron stars. Adding to
this list two pulsar nebulae in the LMC in which pulsars have
been detected, we have thus now
identified the central source in 50\% of SNRs with plerionic
components. The premise that plerions are powered by pulsars therefore
remains essentially unchallenged.

\section*{Observational Properties}

At radio wavelengths, plerions (and plerionic components of composite
SNRs) generally have an amorphous filled-center morphology, with a flat
spectrum ($-0.3 \la \alpha \la 0$, $S_\nu \propto \nu^\alpha$) and a
high degree of linear polarization. In cases where an associated pulsar
has been identified, it is not necessarily near the center and/or
brightest part of the plerion. Kothes \cite{kot98} has proposed that the radio
luminosity of a plerion, $L_R$,  the diameter of the plerion, $D$, and
$\dot{E}$, approximately follow the relation $L_R \propto \dot{E} D$.

\begin{figure} 
\centerline{\epsfig{file=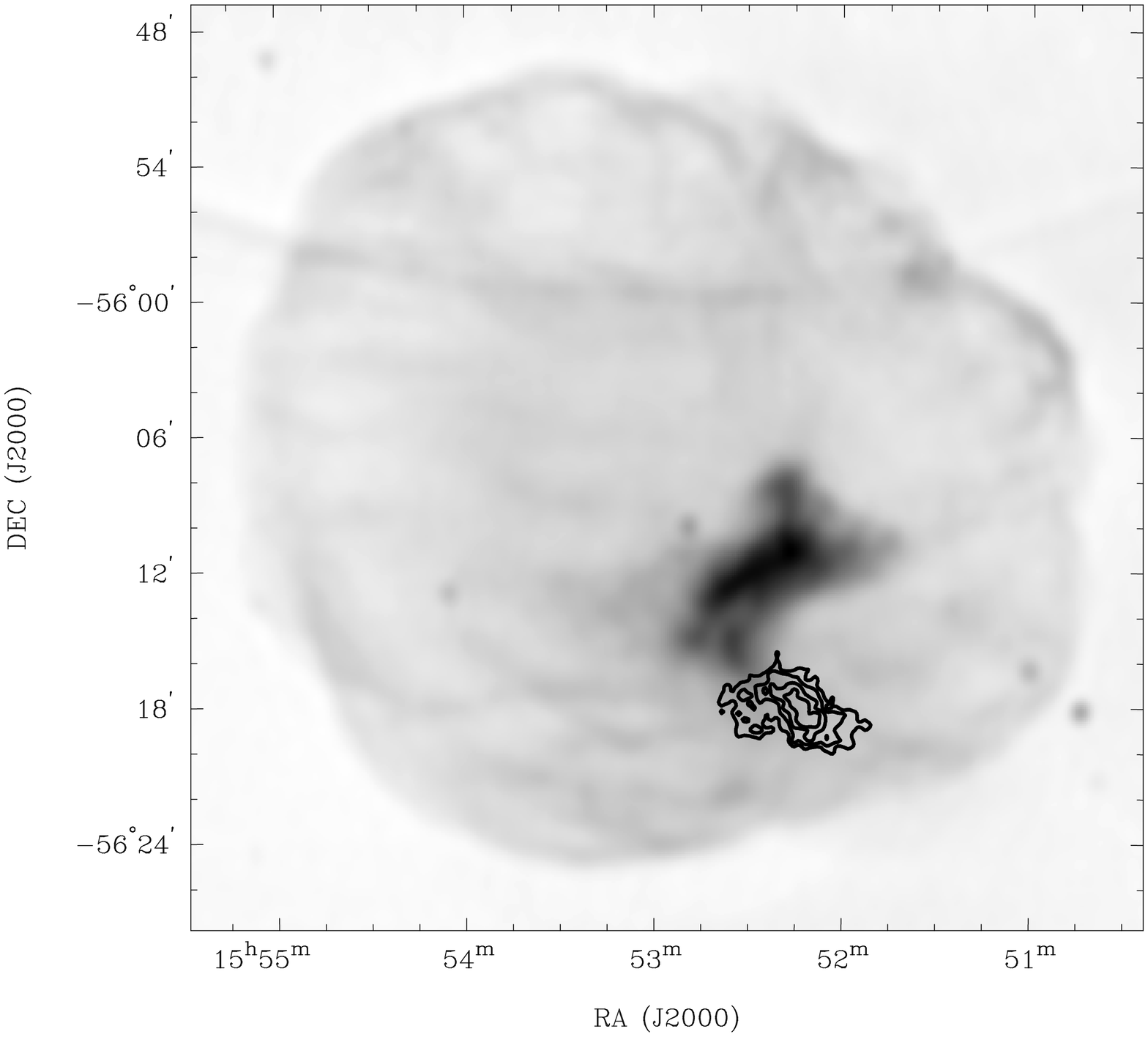,height=8cm,angle=270}
\epsfig{file=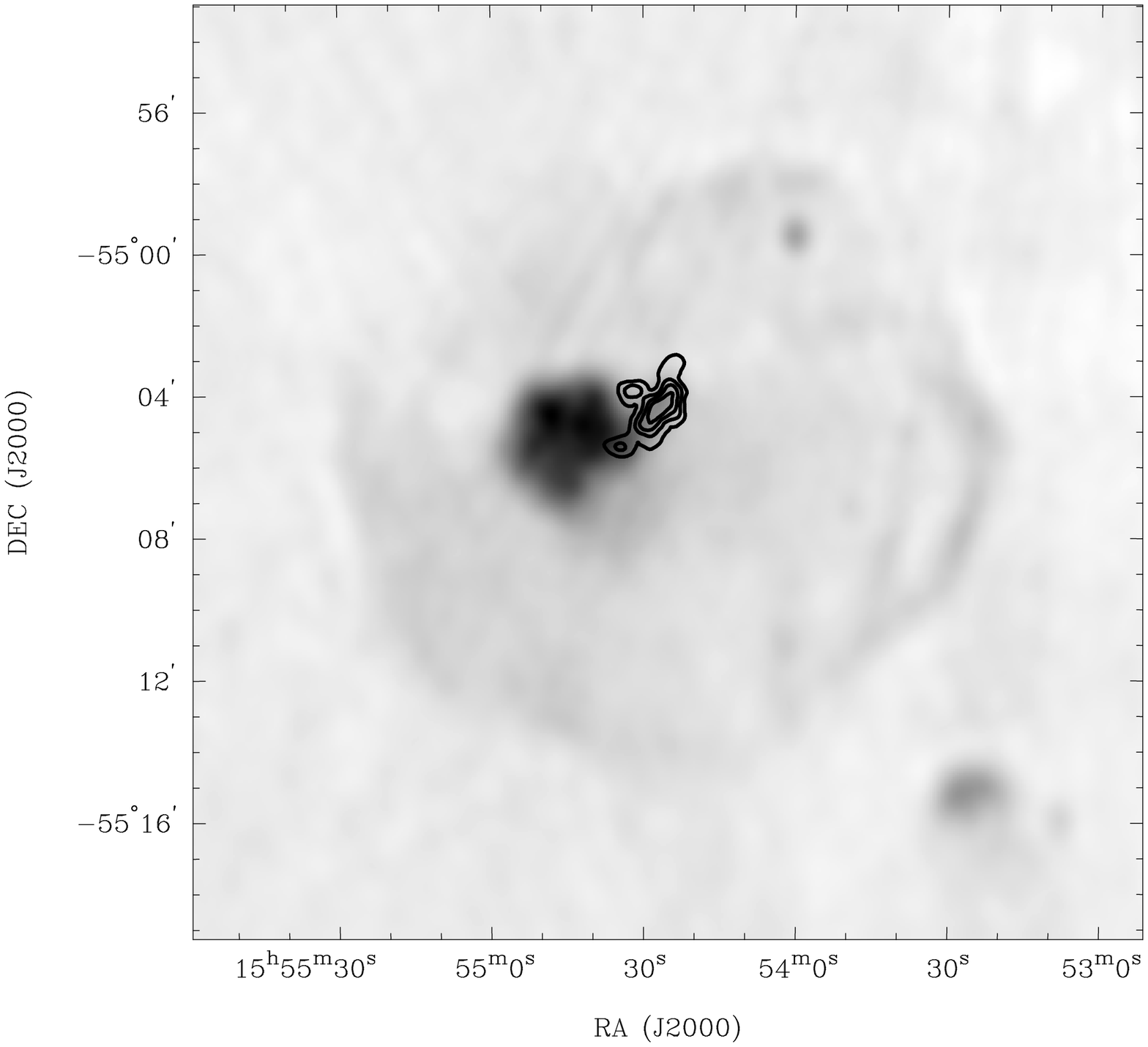,height=8cm,angle=270}}
\caption{The composite SNRs~MSH~15--5{\em 6}\ (left)
and G327.1--1.1 (right). The greyscale shows the
843~MHz radio emission from each remnant \protect\cite{wg96},
demonstrating their composite morphologies. The contours
delineate X-ray emission as seen by {\em ASCA}\ SIS 
\protect\cite{plu98,swc99},
and show the smaller extent and offset of the X-ray
plerions with respect to their radio counterparts.}
\label{fig1}
\end{figure}

In the X-ray regime, plerions imaged at sufficient spatial resolution
seem to contain axially-symmetric structures such as tori and jets.
Plerions generally have smaller X-ray extents than they do in the
radio (see Figure~\ref{fig1}), presumed to be the 
result of the shorter lifetimes of
synchrotron-emitting electrons at progressively higher energies.
For the same reason, 
the spectrum of plerions in X-rays is steeper ($\alpha \sim -1$)
than in the radio, and any associated pulsar
is usually coincident with the brightest X-ray emission. 
Various efforts have been to look for correlations between $\dot{E}$
and the corresponding plerion's X-ray luminosity, $L_X$.
These studies consistently suggest that these two quantities
are correlated, showing that $L_X \propto \dot{E}^a$ with
an exponent in the range $1\la a \la 1.4$ \cite{sw88,bt97}.

\section*{A Plethora of Plerions}

While we assume that all plerions are ``Crab-like'' in that they are 
similarly powered by pulsars, in most respects other plerions have
very different properties to the Crab Nebula. Indeed,
on looking through the available sample,
it becomes apparent that plerions form a very heterogeneous set
of objects. Without wanting to suggest any definitive 
categories\footnote{See Chevalier \cite{che98} 
for a suggested evolutionary sequence for some of these categories.},
some of the different types of object include:

\begin{itemize}
\item ``Standard'' composites, in which the plerionic
component contains a detected pulsar and is surrounded
by a shell. The SNR~0540--69.3 in the LMC \cite{msk93} is a good example
of such a source.

\item ``Naked'' plerions, typified by the Crab Nebula, in
which no surrounding SNR shell can be seen. In the case
of the Crab itself, a good argument can be made that
that a SNR blast-wave is present, but that it
simply can't be seen \cite{sh97}. However, Wallace et al \cite{wlt97,wltp97}
have shown that some of these ``naked'' plerions are interacting
directly with the ambient ISM, and have argued
that such sources are produced by low-energy explosions
in which no fast-moving ejecta are produced. 

\item ``Bow-shock'' plerions such as those
associated with PSRs B1853+01 and B1757--24 \cite{fggd96,fk91}.
These nebulae have a cometary morphology resulting from their pulsar's
high space-velocity. 

\item ``Radio-quiet'' plerions such as 
MSH~15--5{\em 2} and CTA~1 \cite{gbm+98,ssb+97}. In these cases,
a pulsars powers a nebula which is prominent in X-rays but is not seen
at radio wavelengths. It has been proposed that
these plerions are produced by pulsars with high magnetic
fields, $B \ga 10^{13}$~G \cite{bha90,cgk+00}. Such plerions suffer severe
adiabatic losses early in their lives, and so are radio-bright
only at the earliest stages of their evolution.

\item ``Hyper-plerions'', such as G328.4+0.2 and N~157B,
have very large diameters ($\la$20~pc) and radio luminosities
higher than that of the Crab Nebula \cite{gdg00,ldh+00}.
These plerions appear to be powered by
a low field pulsar, $B \la 10^{12}$~G, which
generates a large and long-lived ($\sim$10~kyr) radio nebula \cite{gdg00}.

\item ``Interacting composites'', typified by CTB~80, in
which a high-velocity pulsar catches up with and re-energizes
its associated shell SNR \cite{sfs89}.

\item ``Low frequency spectral break'' plerions, such as 3C~58
and G21.5--0.9. These plerions show a sudden steepening of
their spectra at radio or millimeter frequencies, thought to
be caused by a ``phase change'' in the pulsar's
energy output \cite{wspb97}. These sources will be discussed further below.

\end{itemize}

\section*{Plerion Evolution}

A ``standard'' model of evolution has emerged 
in which a plerion is modeled as a spherically symmetric
expanding synchrotron bubble for which (in the simplest case)
equipartition is assumed between particles and magnetic
fields in the nebula \cite{ps73,rc84}. The only source of energy input
into the nebula is the spin-down of the pulsar,
whose time-evolution is described by:
\begin{equation}
\dot{E} = \frac{L_0}{\left(1+ t/\tau_0\right)^p},~~~p = \frac{n+1}{n-1}
\end{equation}
where $L_0$ is the pulsar's initial spin-down luminosity,
$\tau_0$ is some characteristic time-scale
(typically a few hundred years) and $n$ is 
the pulsar's braking index. For times $t<\tau_0$, the
pulsar's rate of output is approximately constant,
$\dot{E} \approx L_0$. At times $t>\tau_0$,
the spin-down luminosity decays as a power-law, $\dot{E} \propto t^{-p}$.

Competing against this injection
are two sources of energy loss: adiabatic losses due to
expansion of the nebula, and synchrotron losses. Including
all these terms, one can derive expressions for the
evolution of the particle content, magnetic field, luminosity and
spectrum in each phase of evolution. For $t<\tau_0$,
a single break is seen in the plerion's spectrum, corresponding
to the frequency at which synchrotron losses dominate at time $t$.
At times $t>\tau_0$, 
this original spectral break moves to higher frequencies as the
nebular magnetic field decays, while a second ``fossil'' break,
resulting from the rapid decay of $\dot{E}$ beginning
at $t=\tau_0$, appears at lower
frequencies \cite{ps73}.  When modeling the evolution
of a plerion, the presence of a surrounding shell-type SNR,
and/or the actual detection of the associated pulsar, allows
one to estimate the nebular magnetic field strength, the rate of 
energy input by the pulsar and the age of the system.
The properties of the SNR, pulsar and plerion can then
be used to jointly constrain the parameters of the system \cite{sbt98}.

While this picture can explain the basic properties
of the Crab Nebula and other plerions, there have been
many subsequent refinements to take into account particular
situations and
details of the nebular physics. To conclude this section, I
list below some of the recent work that has been carried
out on plerion evolution, and discuss some of the more interesting
developments in more detail.

\begin{itemize}

\item Amato et al \cite{asb+00} have
taken into account the fact that the particle distribution
within a plerion is not homogeneous, and develop a model
in which synchrotron-emitting particles propagate away from
the pulsar.

\item van der Swaluw et al \cite{vag98} have considered the interaction
which occurs when a pulsar catches up with and penetrates
its associated shell SNR.

\item Luz \& Berry \cite{lb99} have modeled the interaction between a
plerion and its surrounding shell SNR.

\item Wilkin \cite{wil00} has developed a detailed treatment of bow-shocks
produced by anisotropic winds, as are likely to be produced
by pulsars.

\item Chevalier \cite{che98} 
has pointed out that the reverse shock produced by
a SNR blast-wave will reach the center of the SNR in $\sim10^4$~yr.
This can compress, brighten and distort a central plerion,
and could account for the filamentary appearance and offset
of the pulsar from the center of the plerion seen in Vela~X.

\item Chevalier \cite{che00} has also recently developed a simplified one-zone
model for the X-ray emission from a plerion.  He derives
an analytic expression for the emission, so that
the X-ray luminosity, $L_X$, depends only on the
spin-down luminosity of the pulsar ($\dot{E}$), the
photon index of the nebula 
($\Gamma$), the wind magnetization parameter ($\sigma$),
the Lorentz factor of the wind ($\gamma_w$) and
the radius of the shock ($r_s$). 
This model successfully predicts the ratio $L_x/\dot{E}$ for
most plerions in which pulsars have been detected.

\item Various authors have considered plerions such
as 3C~58, which have low-frequency
spectral breaks \cite{wspb97,gs92,sbpw98}.  These plerions are
characterized by  sharp spectral breaks ($\Delta\alpha \sim 0.8-1.0$)
at frequencies $\nu_b \la 50$~GHz, in sharp contrast to the
$\Delta\alpha = 0.5$ break seen for the Crab Nebula in the infrared.
The spectral breaks seen in these other plerions cannot be due to
synchrotron losses, as the inferred magnetic field is so high that the
energy in magnetic fields would then be larger than the kinetic energy
of the plerion.  Woltjer et al \cite{wspb97} show that a plerion powered by a
pulsar with a low braking index ($n \ll 3$) can produce a low-frequency
fossil break at times $t > \tau_0$, but that this break is not sharp enough
to match observations. They instead consider a model in which there is
a sudden ``phase change'' in the pulsar's energy output, when perhaps
the pulsar's wind suddenly becomes magnetically dominated. In such a
system a sharp low-frequency spectral break is indeed predicted.  Such
a model can also account for the low X-ray luminosity of 3C~58, and for
the fact that its radio flux is increasing with time (rather than 
decreasing in the case of the Crab). While these arguments make a strong case
that the central sources in these low-frequency break plerions 
are quite different from the Crab Pulsar\footnote{For a different
interpretation in terms of central pulsars with high magnetic
fields, see the discussion by Frail \cite{fra98}.}, the only definitive
resolution to this puzzle will be to actually detect their central
sources.

\end{itemize}

\section*{New Results}

{\bf Gamma rays:}
It has long been thought that many of the unidentified $\gamma$-ray
sources in the Galactic Plane correspond to young high-$\dot{E}$ pulsars 
and their associated plerions. Indeed, Halpern et al \cite{hglh01}
have identified a new radio source, G106.6+3.0,
which is coincident with the otherwise unidentified
{\em EGRET}\ source 3EG~J2227+6122
and also with the X-ray source AX~J2229.0+6114.
This radio source is polarized, has a flat spectral index,
and has a possible bow-shock morphology. The properties
of G106.3+3.0 all 
suggest that this source is a plerion powered
by a pulsar with $\dot{E} \ga 10^{36}$~erg~s$^{-1}$.
Other possible plerions associated with $\gamma$-ray
sources have been reported by Roberts et al \cite{rrjg99}
and Oka et al \cite{okn+99}.

{\bf Radio:} Stappers \& Gaensler have carried
out an extensive search for radio nebulae
associated with radio pulsars \cite{gsfj98,gsf+00,sgj99}. Of 31 pulsars
observed, only one new pulsar nebula was
found, indicating that pulsars reside
predominantly in low-density environments.
Meanwhile, the Australia Telescope Compact
Array continues to image various plerions at
high spatial resolution, highlighting the
diversity of plerion properties and morphologies
\cite{gdg00,ldh+00,dms00}.

{\bf Optical:} The morphology of an optical bow-shock
around a pulsar contains a great deal of information
about a pulsar's interaction with the ISM. Imaging
and spectroscopy of such bow-shocks around
PSRs~B2224+65 (``the Guitar Nebula'') and J0437--4715
have resulted in determinations of
the 3D space velocities of these pulsars, and the
densities and ionization fractions of their environments
\cite{cc00,mrf99}.

{\bf X-rays:} {\em ASCA}\ observations of various pulsars and
their plerions have recently been re-analyzed. Contrary
to previous claims \cite{kt96}, there now seems to be no plerions apparent
around PSRs~B1610--50, B1055--52, B0656+14
or Geminga \cite{sgj99,bkbm99,pkg00}. PSR~B1046--58 has no extended plerion,
but may be associated with a compact X-ray nebula \cite{pkg00}.

With the recent launch of {\em XMM}\ and {\em Chandra}, it
is unsurprising that there are many new results
on pulsars and their nebulae. The high-resolution
of {\em Chandra}\ has been brought to bear on
the two prominent plerions in the Large Magellanic Cloud,
SNRs~0540--693 and N~157B. X-ray data
on the former suggest a possible Crab-like morphology
with the hint of jets and a torus \cite{gw00,kmag+00}, while
observations of the latter appear to confirm the
cometary morphology for this nebula seen in {\em ROSAT} data \cite{tfb+99}.

\begin{figure} 
\centerline{\epsfig{file=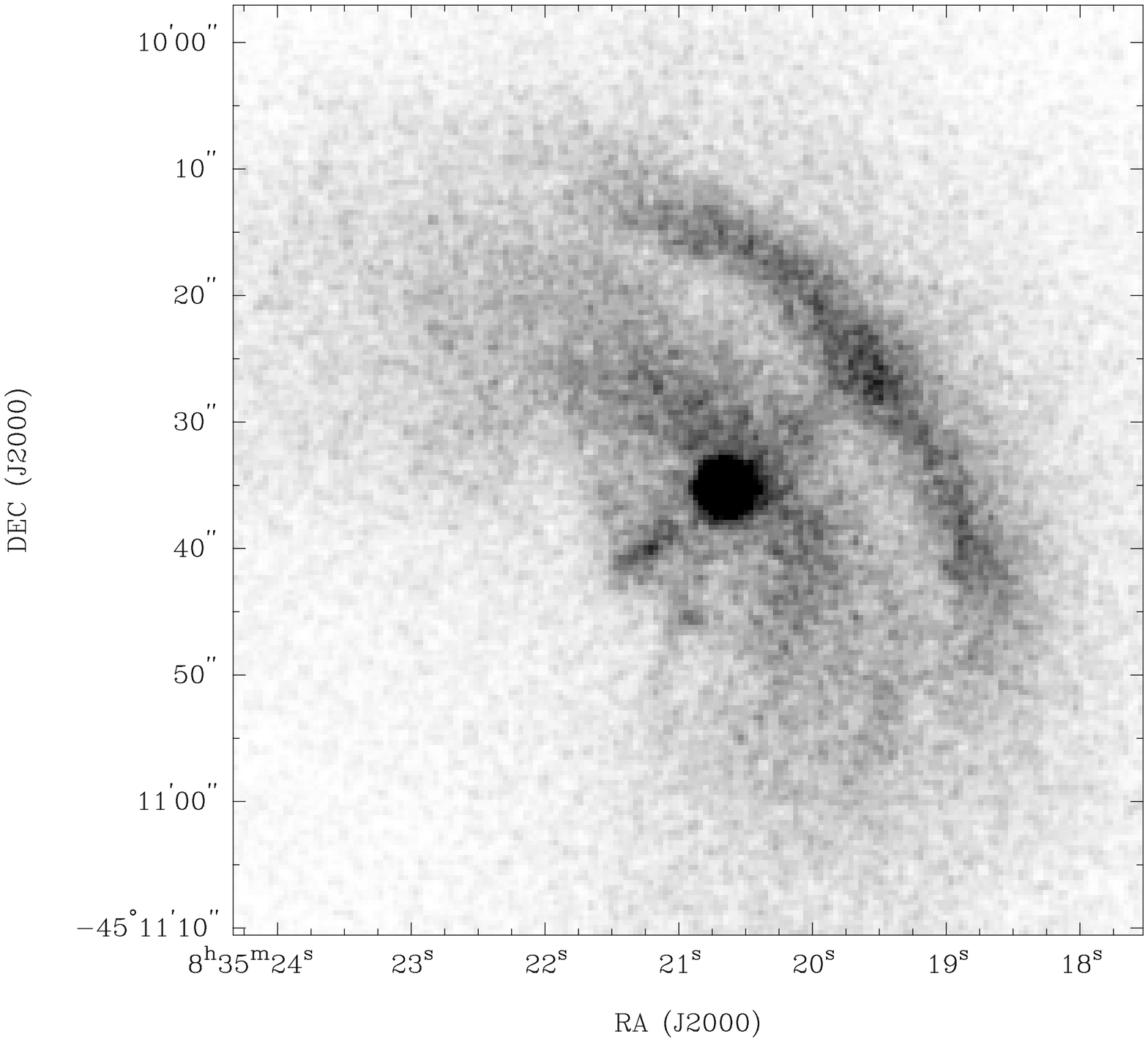,width=8cm,angle=270}
\epsfig{file=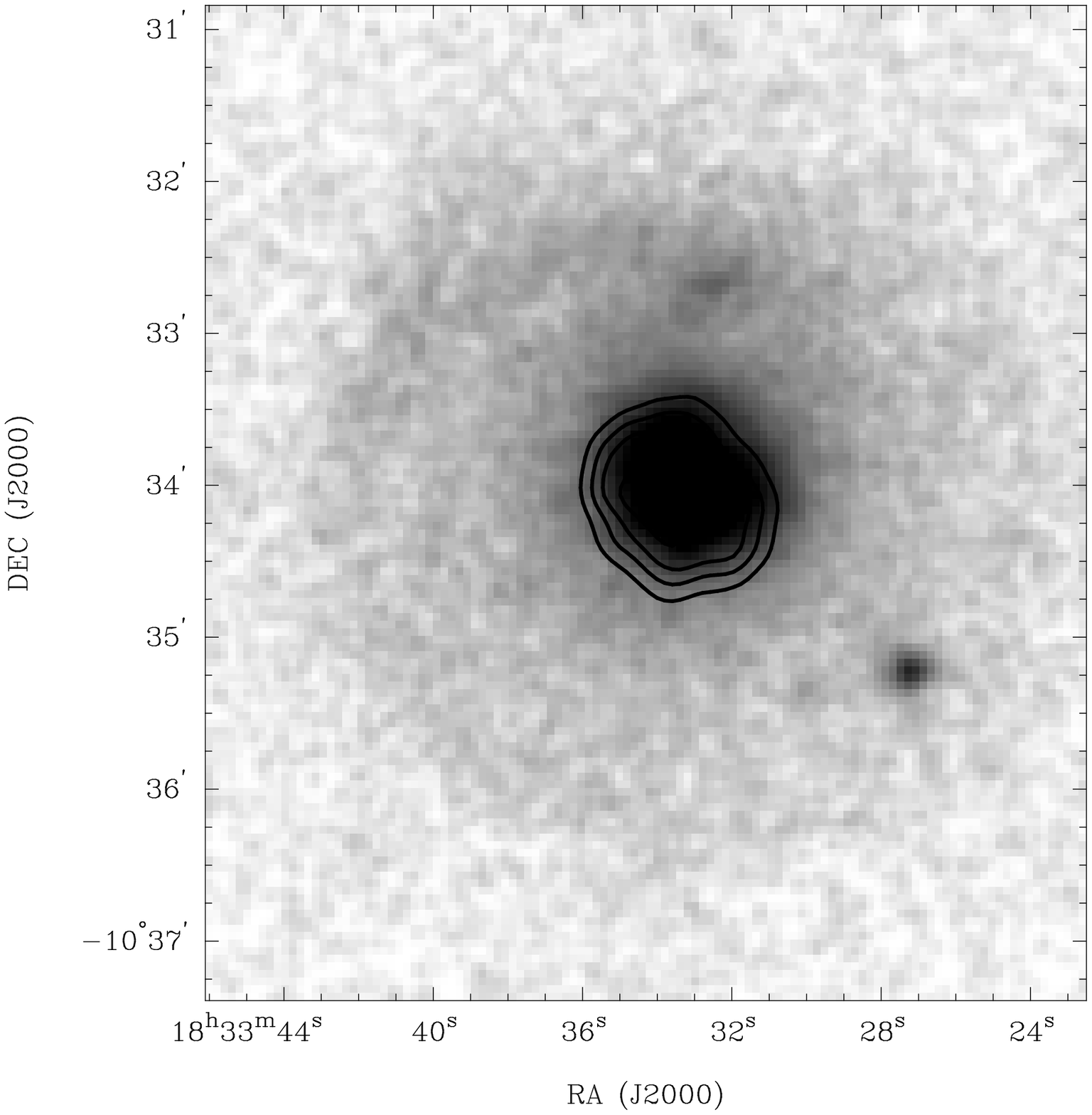,width=8cm,angle=270}}
\caption{The Vela Pulsar and surrounds (left) and SNR~G21.5--0.9 (right).
The image of the Vela Pulsar was produced from archival
{\em Chandra}\ HRC data, and has been
convolved with a $0\farcs5$ gaussian (see also \protect\cite{hgh01}).
G21.5--0.9 was observed with {\em XMM}\ EPIC MOS \protect\cite{wbb+01} ---
the contours show 1.4~GHz radio emission from the plerion
indicating its previously-known extent, while
the greyscale has been scaled logarithmically to show the faint
surrounding X-ray halo.}
\label{fig2}
\end{figure}

The {\em Chandra}\ image of the Vela Pulsar
is spectacular (Figure~\ref{fig2}; \cite{hgh01}). The surrounding
nebula is remarkably similar to the Crab,
showing clear evidence for equatorial rings
and axial jets, and  appearing to rule out
earlier interpretations of this system 
as a bow-shock. The orientation of the pulsar's proper motion,
the pulsar's spin axis and the direction
of outflow along the X-ray jets all seem to align.
This alignment is similar to that seen for
the Crab Pulsar, and provides important
constraints on the origins of pulsar
spin periods and space velocities \cite{lcc01,sp98}.
Helfand et al \cite{hgh01} note that
the X-ray properties of the Vela plerion 
suggest a magnetization parameter
$\sigma \approx 1$, in sharp contrast
to the Crab Nebula for which $\sigma \approx 3 \times 10^{-3}$.

Finally, {\em Chandra}\ and {\em XMM}\ observations
of the plerion G21.5--0.9 have revealed both a
compact central source and a surrounding halo 
(Figure~\ref{fig2}).
The central source is resolved by {\em Chandra},
and may be a wisp or termination shock which
is hiding the pulsar itself \cite{scs+00}. Meanwhile,
{\em XMM}\ observations clearly demonstrate
that the surrounding halo has a power-law spectrum,
and that this spectrum steadily steepens with radius \cite{wbb+01}. 
It is not yet clear whether the plerion 
is much larger than previously thought, or if
the halo is a surrounding SNR blast-wave so that
G21.5--0.9 should be re-classified as a composite SNR.
Deep radio observations of this
source will be required to distinguish between these possibilities. 

\section*{Conclusions}

With the high-quality data now available across the spectrum,
it has become abundantly apparent that the Crab Nebula is not typical
of plerions. The large diversity in the observed
and inferred properties of plerions appears to result
from differences in their environments, ages
and associated pulsars. Key to understanding 
this variety seems to be better modeling of the
interaction of plerions and pulsars with their surrounding
SNRs. 

It is also clear that many plerions are still waiting
to be discovered. Approximately 20\% of Galactic radio
SNRs are yet to be imaged with a spatial resolution  of
better than $\sim10$~beams across their diameter, and imaging
these SNRs at higher resolution could reveal central
pulsar-powered components. Furthermore, many pulsar
nebula are seen in hard X-rays but not at radio wavelengths,
and so many apparently shell-type SNRs which have not yet been observed
at higher energies may also harbor a central plerion.

A glance at the list of targets approved for {\em Chandra}\
and {\em XMM}\ shows that many plerions have been or are
about to be observed with these new instruments. These data
will obviously produce a great deal more information on some
of the issues touched on here. Do all pulsar winds 
show a ``torus $+$ jets'' morphology? Is the alignment
between rotation axis and proper motion seen for the Crab
and Vela pulsars a common characteristic? Do anomalous
X-ray pulsars and soft $\gamma$-ray pulsars power associated
nebula? Do ``naked'' plerions have
faint surrounding shells? We can look forward to 
a whole new picture of plerions emerging in the near future. \\

%%%% See AIP examples for more details on insertion of figures and tables.

%\section*{Acknowledgments} 

I thank Lisa Townsley and Jasmina Lazendic for sharing
with me their recent work on N~157B,
Bob Warwick for providing an {\em XMM}\ image of G21.5--0.9,
and all those with whom I work on plerions and pulsars
for many interesting discussions and collaborations.
This work was supported by NASA through Hubble
Fellowship grant HST-HF-01107.01-A awarded by the Space Telescope
Science Institute, and by NSF under Grant No. PHY99--07949.

%\bibliographystyle{aipproc}
%\bibliography{journals,modrefs,psrrefs,crossrefs}

\begin{thebibliography}{10}

\bibitem{sbt98}
Slane, P., Bandiera, R., and Torii, K., 1998, Mem. Soc. Astron. It., {\rm
69}, 945

\bibitem{che98}
Chevalier, R.~A., 1998, Mem. Soc. Astron. It., {\rm 69}, 977

\bibitem{fra98b}
Frail, D.~A., in \emph{Neutron Stars and Pulsars: Thirty Years after the
Discovery}, edited by N.~Shibazaki, N.~Kawai, S.~Shibata, and T.~Kifune,
Tokyo: Universal Academy Press, 1998 p.~423

\bibitem{fra98}
Frail, D.~A., in \emph{The many faces of neutron stars}, edited by
R.~Buccheri, J.~van Paradijs, and M.~A. Alpar, vol. 515 of 
\emph{NATO ASI Series}, Dordrecht: Kluwer, 1998 p.~179

\bibitem{gae01}
Gaensler, B.~M., in \emph{Spin, Magnetism and Cooling of Young Neutron
Stars}, Santa Barbara: Institute for Theoretical Physics, 2000
    (http://online.itp.ucsb.edu/online/neustars\_c00/gaensler)

\bibitem{lun21}
Lundmark, K., 1921, Publ. Astr. Soc. Pacific, {\rm 33}, 225

\bibitem{sr68}
Staelin, D.~H., and Reifenstein, {III}, E.~C., 1968, Science, {\rm 162}, 1481

\bibitem{ws71}
Weiler, K.~W., and Seielstad, G.~A., 1971, ApJ, {\rm 163}, 455

\bibitem{lgcm77}
Lockhart, I.~A., Goss, W.~M., Caswell, J.~L., and McAdam, W.~B., 1977, MNRAS,
 {\rm 179}, 147

\bibitem{wp78}
Weiler, K.~W., and Panagia, N., 1978, A\&A, {\rm 70}, 419

\bibitem{ws78}
Weiler, K.~W., and Shaver, P.~A., 1978, A\&A, {\rm 70}, 389

\bibitem{sha79}
Shakeshaft, J.~R., 1979, A\&A, {\rm 72}, L9

\bibitem{mgh+79}
Milne, D.~K., et al, 1979, MNRAS, {\rm 188}, 437

\bibitem{gre00}
Green, D.~A., \emph{A {C}atalogue of {G}alactic {S}upernova {R}emnants (2000
  {A}ugust {V}ersion)}, Cambridge: Mullard Radio Astronomy Observatory, 2000.
   (http://www.mrao.cam.ac.uk/surveys/snrs/)

\bibitem{kot98}
Kothes, R., 1998, Mem. Soc. Astron. It., {\rm 69}, 971

\bibitem{wg96}
Whiteoak, J. B.~Z., and Green, A.~J., 1996, A\&AS, {\rm 118}, 329,
  (http://www.physics.usyd.edu.au/astrop/wg96cat/)

\bibitem{plu98}
Plucinsky, P.~P., 1998, Mem. Soc. Astron. It., {\rm 69}, 939

\bibitem{swc99}
Sun, M., Wang, Z., and Chen, Y., 1999, ApJ, {\rm 511}, 274

\bibitem{sw88}
Seward, F.~D., and Wang, Z.-U., 1988, ApJ, {\rm 332}, 199

\bibitem{bt97}
Becker, W., and Tr\"{umper}, J., 1997, A\&A, {\rm 326}, 682

\bibitem{msk93}
Manchester, R.~N., Staveley-Smith, L., and Kesteven, M.~J., 1993, ApJ, {\rm
  411}, 756

\bibitem{sh97}
Sankrit, R., and Hester, J.~J., 1997, ApJ, {\rm 491}, 796

\bibitem{wlt97}
Wallace, B.~J., Landecker, T.~L., and Taylor, A.~R., 1997, AJ, {\rm 114},
2068

\bibitem{wltp97}
Wallace, B.~J., Landecker, T.~L., Taylor, A.~R., and Pineault, S., 1997,
A\&A, {\rm 317}, 212

\bibitem{fggd96}
{Frail}, D.~A., {Giacani}, E.~B., {Goss}, W.~M., and {Dubner}, G., 1996, ApJ,
{\rm 464}, L165

\bibitem{fk91}
Frail, D.~A., and Kulkarni, S.~R., 1991, Nature, {\rm 352}, 785

\bibitem{gbm+98}
Gaensler, B.~M., et al, 1999, MNRAS, {\rm 305}, 724

\bibitem{ssb+97}
Slane, P., et al, 1997, ApJ,
  {\rm 485}, 221

\bibitem{bha90}
Bhattacharya, D., 1990, J. Astrophys. Astr., {\rm 11}, 125

\bibitem{cgk+00}
Crawford, F., et al, 2001, ApJ, in press (astro-ph/0012287)

\bibitem{gdg00}
Gaensler, B.~M., Dickel, J.~R., and Green, A.~J., 2000, ApJ, {\rm 542}, 380

\bibitem{ldh+00}
Lazendic, J.~S., et al, 2000, ApJ, {\rm 540}, 808

\bibitem{sfs89}
Shull, J.~M., Fesen, R.~A., and Saken, J.~M., 1989, ApJ, {\rm 346}, 860

\bibitem{wspb97}
Woltjer, L., Salvati, M., Pacini, F., and Bandiera, R., 1997, A\&A, {\rm
325}, 295

\bibitem{ps73}
Pacini, F., and Salvati, M., 1973, ApJ, {\rm 186}, 249

\bibitem{rc84}
Reynolds, S.~P., and Chevalier, R.~A., 1984, ApJ, {\rm 278}, 630

\bibitem{asb+00}
Amato, E., et al, 2000, A\&A,
  {\rm 359}, 1107

\bibitem{vag98}
van~der Swaluw, E., Achterberg, A., and Gallant, Y.~A., 1998, Mem. Soc.
Astron.  It., {\rm 69}, 1017

\bibitem{lb99}
Luz, D. M. G.~C., and Berry, D.~L., 1999, MNRAS, {\rm 306}, 191

\bibitem{wil00}
Wilkin, F.~P., 2000, ApJ, {\rm 532}, 400

\bibitem{che00}
Chevalier, R.~A., 2000, ApJ, {\rm 439}, L45

\bibitem{gs92}
Green, D.~A., and Scheuer, P. A.~G., 1992, MNRAS, {\rm 258}, 833

\bibitem{sbpw98}
Salvati, M., Bandiera, R., Pacini, F., and Woltjer, L., 1998, Mem. Soc.
Astron.  It., {\rm 69}, 1023

\bibitem{hglh01}
Halpern, J.~P., Gotthelf, E.~V., Leighly, K.~M., and Helfand, D.~J., 2001,
ApJ, in press (astro-ph/0007076)

\bibitem{rrjg99}
Roberts, M. S.~E., Romani, R.~W., Johnston, S., and Green, A.~J., 1999, ApJ,
{\rm 515}, 712

\bibitem{okn+99}
{Oka}, T., et al, 1999, ApJ, {\rm 526}, 764

\bibitem{gsfj98}
Gaensler, B.~M., Stappers, B.~W., Frail, D.~A., and Johnston, S., 1998, ApJ,
{\rm 499}, L69

\bibitem{gsf+00}
Gaensler, B.~M., et al, 2000, MNRAS, {\rm 318}, 58

\bibitem{sgj99}
Stappers, B.~W., Gaensler, B.~M., and Johnston, S., 1999, MNRAS, {\rm 308},
609

\bibitem{dms00}
Dickel, J.~R., Milne, D.~K., and Strom, R.~G., 2000, ApJ, {\rm 543}, 840

\bibitem{cc00}
Chatterjee, C., and Cordes, J.~M., in \emph{Pulsar Astronomy --- 2000 and
  Beyond, {IAU} Colloquium 177}, edited by M.~Kramer, N.~Wex, and
    R.~Wielebinski, San Francisco: Astronomical Society of the Pacific, 2000
      p.~517

\bibitem{mrf99}
Mann, E.~C., Romani, R.~W., and Fruchter, A.~S., 1999, BAAS, {\rm 195}, 41.01

\bibitem{kt96}
Kawai, N., and Tamura, K., in \emph{Pulsars: Problems and Progress, {IAU}
  Colloquium 160}, edited by S.~Johnston, M.~A. Walker, and M.~Bailes, San
    Francisco: Astronomical Society of the Pacific, 1996 p.~367

\bibitem{bkbm99}
Becker, W., Kawai, N., Brinkmann, W., and Mignani, R., 1999, A\&A, {\rm 352},
 532

\bibitem{pkg00}
Pivovaroff, M., Kaspi, V.~M., and Gotthelf, E.~V., 2000, ApJ, {\rm 528}, 436

\bibitem{gw00}
Gotthelf, E.~V., and Wang, Q.~D., 2000, ApJ, {\rm 532}, L117

\bibitem{kmag+00}
Kaaret, P., et al, 2000, ApJ, in press (astro-ph/0008388)

\bibitem{tfb+99}
Townsley, L., et al, 1999, BAAS, {\rm 195}, 53.01

\bibitem{hgh01}
Helfand, D.~J., Gotthelf, E.~V., and Halpern, J.~P., 2001, ApJ,
  submitted (astro-ph/0007310)

\bibitem{lcc01}
Lai, D., Chernoff, D.~F., and Cordes, J.~M., 2001, ApJ, {\rm 549}, in press
  (astro-ph/0007272)

\bibitem{sp98}
Spruit, H., and Phinney, E.~S., 1998, Nature, {\rm 393}, 139

\bibitem{scs+00}
Slane, P., et al, 2000, ApJ, {\rm 533}, L29

\bibitem{wbb+01}
Warwick, R.~S., et al, 2001,
  A\&A, in press (astro-ph/0011245)

\end{thebibliography}

\end{document}